\documentclass[12pt]{article}

\usepackage{amsmath}				
\usepackage{amssymb}				
\usepackage{amsfonts,commath,bm} 	
\usepackage{setspace}				
\usepackage{graphicx}				
\usepackage{wrapfig}				
\usepackage{times}					

\usepackage[top=1in, bottom=1in, left=1in, right=1in]{geometry} 	
\usepackage[compact]{titlesec}								
\usepackage[font=footnotesize,labelfont=bf,labelsep=period,width=0.75\textwidth]{caption} 	
\usepackage[font=footnotesize,labelfont=bf,labelsep=period]{subcaption} 				
\DeclareCaptionSubType*[arabic]{figure} 										
\DeclareCaptionLabelFormat{subfiglabel}{Figure #2} 								
\usepackage[capitalize]{cleveref}
\crefname{figure}{Fig.}{Figs.}
\Crefname{figure}{Figure}{Figures}
\crefname{table}{Tab.}{Tabs.}
\Crefname{table}{Table}{Tables}
\crefname{equation}{Eq.}{Eqs.}
\Crefname{equation}{Equation}{Equations}
\crefname{section}{Sec.}{Secs.}
\Crefname{section}{Section}{Sections}

\usepackage{ifthen}
\usepackage[dvipsnames]{xcolor}
\usepackage[normalem]{ulem}

\begin{document}

\begin{center}

{\Large Rubber Elasticity: Solution of the James-Guth Model}\par\vskip12pt

{\normalsize B. E. Eichinger}\par
    
{\small\itshape Department of Chemistry, University of Washington, Seattle, Washington 98195-1700}\par


\end{center}

\begin{abstract}
The solution of the many-body statistical mechanical theory of elasticity formulated by James
and Guth in the 1940s is presented. The remarkable aspect of the solution is that it gives an elastic free energy that is essentially equivalent to that developed by Flory over a period of several decades.
\end{abstract}
\section*{Introduction}
Rubber elasticity is the first bulk property of polymers that yielded to theoretical analysis. The identification of the relation between the Gaussian distribution of the end-to-end distance of a random walk with the quadratic strain dependence of phenomenological stress-strain theory was key to this success. Of course, the underlying physics that makes this connection inevitable and viable is that a polymer chain in the bulk amorphous ÒmeltÓ phase is unperturbed by intermolecular interactions.\cite{1} This fact allows one to realize the force that a chain delivers to the cross-links that terminate it, and which tie it to other chains in the three-dimensional space filling random network, is determined solely by the chain's intramolecular potential. While high elasticity theory was the first to successfully predict bulk polymeric materials behavior, it remains one of the few, and perhaps the only, analytical theory of polymers to do so, computer simulations notwithstanding. This is reason enough to justify efforts to improve upon the theory.  Given the history of the subject, rubber elasticity is one of the first soft materials to admit an atom-based theoretical analysis.

The extent to which the elastic equation of state is determined by the interaction between network connectivity and chain statistics has been a point of contention from the earliest days of polymer theory. The theory initiated by Kuhn,\cite{2} elaborated by Wall\cite{3,4}and Flory and Rehner,\cite{5,6} and discussed extensively in treatises,\cite{1,7,8,9,10} is constructed by adding together the contributions to the stress from independent chains. This requires the so-called affine assumption -- the displacement of the ends of an average network chain is congruent to the macroscopic strain. This theory is relatively easy to execute, but by treating the chains as independent it incurred the criticism of James and Guth.\cite{11,12,13} In their many-body theory the individual chains obey Gaussian statistics, just as in the independent chain theory, but James and Guth emphasized the fact that in tying the chains together with cross-links they become an indissoluble whole that must be treated as a single entity. This insight carried a heavy price -- their many-body formulation was too difficult to be convincingly solved. To make progress, James and Guth\cite{11} introduced the unphysical notion of fixed junctions (fixed by the external constraints) that are displaced by the macroscopic strain. However, the value that one deduces for the modulus depends on the choice and number of fixed junctions,\cite{14} which effectively ruins any rigor that may be ascribed to the many-body theory.

Improvements to the independent chain theory by Flory and his colleagues over the span of about forty years introduced successive improvements to the theory. In treating swelling, Flory\cite{15} introduced the controversial combinatorial term giving a contribution $-(2\nu k/f)\ln (V)$ to the entropy. Here $\nu$ is the number of chains in the network of volume $V$, $f$ is the functionality of the cross-links, and $k$ is Boltzmann's constant. Using somewhat convoluted reasoning,\cite{16,17} the ratio $\langle r_i^2\rangle/ \langle r^2\rangle_0$ was inserted into the theoretical Young's modulus in the 1950s. The numerator in this expression is the value of the mean-square end-to-end distance of the average network chain in a reference state and the denominator is the similar quantity for the free unperturbed chain at temperature $T$ . By involving chain dimensions in the modulus, Flory and coworkers were able to evaluate the temperature dependence of chain dimensions in terms of the stress-temperature coefficient. (It will be noted that the unperturbed chain dimensions drop out of the modulus in the independent chain theory, so an additional argument is needed to put this term into the modulus.) This ratio of dimensions was not featured in Flory's work\cite{14} of the 1970s, where the number of chains is replaced by the cycle rank to account for the contribution from cross-links that are inserted after the gel point. This appears to have been motivated by the idea that up to the gel point the nascent network, which is approximated as an acyclic tree, cannot support an equilibrium stress.  Given this idea, it is only the cross-links that are inserted after the gel point that contribute to the stress.  Assuming that all the prepolymer has been incorporated into the tree at the gel point, the cross-links that are inserted subsequently can only form cycles in the tree. The cycle rank measures the number of chains that are ``activated" by these post-gel cross-links.  In replacing the number of chains by the cycle rank, the modulus decreases to a value close to that advocated by James and Guth.

High elasticity has attracted the attention of theorists far too numerous to mention here. Much of this work has been reviewed in articles\cite{18,19} and monographs.\cite{9,10} This paper is not intended to be a comprehensive review of all the important work that has gone into our current understanding of the physics and chemistry of elastomers and gels. It is instead aimed at solving the James-Guth many-body theory in the small strain limit. The objective of the paper is to eliminate as many physical assumptions as possible beyond those that comprise the basic model, and to make mathematical approximations clear so that appropriate confirmation or improvements might be made in the future.  The remarkable result of this calculation is that it gives all the terms that Flory put in by hand over the years, although the interpretation of the terms is somewhat different. There is no essential difference between the James-Guth and Flory-Wall theories!  The other aspect of this work to emphasize is that it firmly establishes a baseline against which one may make quantitative statements about other models, including those that treat entanglements, {\emph e.g.}, tube models. 

\section*{Configuration Integral and Potential of Mean Force}
The shape of a soft system is a characteristic of fundamental interest, and unlike a gas or liquid, where the volume is fixed by a container, a soft system adopts a macroscopic shape that is determined by a combination of internal potentials and external constraints that are imposed at the time of formation.  Once formed, an unconstrained soft system is free to adjust its boundary based only on internal potentials. This section formalizes the computation of the probability function that is required to specify the geometric information that characterises the size and shape of the body of interest.

The probability that a classical system described by the Hamiltonian $H(p,q)$, and having thermal energy $kT$, is found in a state $\{p,q\}$, where $p$ and $q$ are $3N$ dimensional momenta and coordinate vectors, respectively, is proportional to $\exp[-H(p,q)/ kT]$. The $p$ and $q$ of a classical system are continuously variable, such that the probability $P(p,q)dpdq$ is defined in the $6N$-dimensional coordinate patch $dpdq$ by
\begin{equation*}
P(p,q)dpdq = Z^{-1}\exp[-H (p,q)/kT]dpdq
\end{equation*}
where $Z$ normalizes the distribution. Now define a function ${\bf F}(p,q)$, which in general can be a tensor valued function of any degree, with corresponding volume element $d{\bf F}$. The probability $P({\bf F})d{\bf F}$ that the system will be found in a state with value ${\bf F}$ is given by
\begin{equation*}
P({\bf F})d{\bf F}=d{\bf F}\int P(p,q)dpdq/d{\bf F},
\end{equation*}
with the integration being performed over the space complementary to ${\bf F}$, which is the meaning of $dpdq/d{\bf F}$.  

The change in free energy, $\Delta A$, accompanying a change in state of the system from ${\bf F}_1$ to ${\bf F}_2$ is given by
\begin{equation}\label{eq1}
\Delta A=-kT\ln [P({\bf F}_2 )d{\bf F}_2/P({\bf F}_1)d{\bf F}_1]=-kT\ln[J(¶{\bf F}_2/¶{\bf F}_1)P({\bf F}_2)/ P({\bf F}_1)]
\end{equation}
where $d{\bf F}_2/d{\bf F}_1=J(¶{\bf F}_2/¶{\bf F}_1)$ is the Jacobian determinant of the mapping ${\bf F}_1\to {\bf F}_2$ . In most cases the Jacobian determinant will not be commensurate with $P({\bf F})$ and will therefore not contribute to the thermodynamics. The reversible work done on the system to convert it from state ${\bf F}_1$ to state ${\bf F}_2$ is just $\Delta A$. Because it is only the ratio of probabilities that is important in thermodynamics, the normalization by $Z$ is immaterial and will be dropped where no harm comes from doing so. 

The elastic body of interest consists of a single covalently bonded gel component, with no sol fraction. For this classical problem the momenta integrate trivially and one is left with the calculation of the unnormalized probability distribution
\begin{equation}\label{eq2}
P({\bf S},T)d{\bf S} = d{\bf S} \int \exp[-\beta V (q)]dq/d{\bf S}
\end{equation}
that the system consisting of $N$ (implicit) polymer atoms at temperature $T$ has a shape and size determined by ${\bf S} = {\bf S}(q)$ (to be specified later). Here $V (q)$ is the potential energy of the system, $\beta = 1/kT$ , and $dq$ is the $3N$-dimensional volume element. (At this early stage swelling can be formally accommodated with use of a semi-grand ensemble for a polymer-solvent system -- the system can be open with respect to exchange of solvent molecules at specified chemical potential. This paper focuses on the stress-stain relation and not on swelling, so that elaboration is not pursued.)

The ``moding-out" operation, $dpdq\to dpdq/d{\bf F}$, is now performed once more, but on the inside of the shape distribution. Let there be $\mu$ cross-link ``atoms", and consider performing the integral in eq. (2) in two steps as
\begin{equation*}
P({\bf S},T)d{\bf S}=d{\bf S}\int dq_{\mu}\int \exp[-\beta V (q)]dq/dq_{\mu}d{\bf S}. 
\end{equation*}
That is, first integrate over the mid-chain coordinates between the cross-links while holding the latter coordinates fixed, after which the integrations over the cross-link positions are to be executed.  The first integration generates an acceptable approximation to the free energy for the un-cross-linked polymer together with a remaining piece that is determined by the potential of mean force acting between the cross-links. That is
\begin{equation}\label{eq3}
P({\bf S},T)d{\bf S}=\exp[-\beta A_0({\bf S},T)]d{\bf S}\int\exp[-\beta \bar V(q_{\mu})]dq_{\mu}/d{\bf S}
\end{equation}
where $A_0({\bf S},T)$ is the free energy of the un-cross-linked polymer.  This function is presumed to depend
only on the volume of the elastomer and not on its shape; this will be defined more precisely after ${\bf S}$ is specified. What remains under the integral is the Boltzmann factor of the potential of mean force, $\bar V(q_{\mu})$, acting between cross-links. [The Flory-Rehner theory of swelling\cite{5,6} assumes this separation of the free energy of the base polymer from that of the elastomer.  Sensitive swelling experiments\cite{6a,6b,6c} suggest that this separation is not strictly valid.  A theory that couples cross-link modes of motion with mid-chain motions would take us far afield of the present objective, which is restricted to an analysis of the classical theory and the presentation of techniques for handling soft materials.]

The potential of mean force, $\bar V(q_{\mu})$ , for an elastomer is given in all elementary elasticity theories
as a product of Boltzmann factors, one for each chain in the network. James and Guth\cite{11} wrote this product as a many-body quadratic potential of the form
\begin{equation}\label{eq4}
\bar V/kT =\sum_{i-j}\gamma_{ij}(r_i-r_j)^2=\gamma {\rm tr}(XKX') 
\end{equation}
where $\gamma_{ij}= 3/2\langle r^2\rangle_{ij}$ is the modulus parameter for the tie-chains, with $\langle r^2\rangle_{ij}$ being the mean-square displacement, at the implicit temperature $T$, between junction pairs $i-j$ that are directly connected by a single chain. (James and Guth formulated the potential with a distribution of chain lengths, as has been done here. The simplified, second version in eq. (4) is written with an \emph {average} $\gamma$ to emphasize the chain dimension parameter that will carry important information.  For a network constructed with uniform chains this is the sole required molecular characteristic; for a distribution of chain lengths the averaging over the chain length distribution will be addressed later.) The Cartesian coordinates of the junction points are written as the $3\times \mu$ matrix $X = (x^a_i); 1 \le a\le 3, 1\le i \le \mu$, with transpose $X'$. There may be multiple chains that directly connect two junctions, but this is incorporated in the matrix $K$, which is the Laplacian for the graph\cite{22} that encodes the connectivity of the network. (In previous work the author\cite{23} named this the Kirchhoff matrix. The mathematical literature\cite{22} has settled on the name Laplacian. It would also be appropriate to call this the Hessian for the network.) Regardless of the name, the construction of the matrix has been described in detail in several publications\cite{11,23,24} and need not be repeated here. It is assumed that the Laplacian has a single zero eigenvalue, which signifies that the system consists of a single connected component.

\section*{Brout-Fixman-Edwards (BFE) Averaging}

The next formality turns out to be extremely important for practical evaluation of the stress-strain relation: we need to average over frozen disorder. Systems with frozen disorder require a higher level average of the free energy than is usually encountered in elementary statistical mechanics. This was first described by Brout\cite{25} in a treatment of order-disorder transitions; Fixman used the averaging in the guise of conditional probabilities in work on polypeptides\cite{26} and polynucleic acids;\cite{27} and finally, Edwards and coworkers\cite{28} formalized the averaging in several treatments of rubber elasticity. Let $\{C\}$ be a set of internal constraints. The set $\{C_i\}$ is a particular instantiation of the constraints that characterizes a member of the ensemble of samples of the material, all members having been prepared under the action of identical external constraints. The probability that a system is observed with this set of constraints is $P(T_* ,\{C_i\})\propto \exp[-A(T_*,\{C_i\})/kT_*]$, where $A(T_*,\{C_i\})$ is the free energy of the system that is formed at temperature $T_*$ when the constraints $\{C\}$ are imposed. While the constraints are formed at temperature $T_*$, they remain fixed when the system temperature is changed to $T$. The BFE average that we need is
\begin{equation}\label{eq5}
\langle A(T)\rangle= \sum_i P(T_*,\{C_i\}) A(T,\{C_i\}) 
\end{equation}
The average $\langle A(T)\rangle$ depends on additional parameters that are implicit. For the elasticity problem, the cure temperature is $T_*$, and the constraints are the cross-linkages that are formed at cure.

\section*{The Holonomic Constraint Trick}
The exact stress-strain relation from continuum mechanics\cite{29} is 
\begin{equation}\label{murg}
\sigma=2(\rho_{\varepsilon}/\rho_1)\varepsilon[\partial a/\partial(\varepsilon'\varepsilon)]_T\varepsilon'
\end{equation}
where $\sigma$ is the stress tensor, $\rho_{\varepsilon}$ is the mass density at the state of strain specified by the deformation gradient tensor $\varepsilon$ (for the unstrained state, $\varepsilon =1=\textrm{unit tensor}$), $\varepsilon'$ is the transpose of $\varepsilon$, and $a$ is the Helmholtz free energy per unit volume in the unstrained state. All of these quantities are defined point-wise, and in general will vary from point to point in the medium.  While the equation is an exact continuum equation, it is untenable from the standpoint of statistical mechanics. Suppose that one has an inhomogeneous medium in which the phases are sufficiently finely dispersed that the interaction between neighboring phases is a substantial contribution to the free energy density. How does one evaluate the free energy density for a domain? It is not the discontinuities per se that cause problems -- it is the interactions between neighboring volume elements that defeat evaluation of a free energy density. The potential energy of atoms that interact across bounding surfaces of volume elements cannot be assigned unequivocally to the elements on either side of the surface. This is one of the more glaring examples of the incompatibility of continuum and atomic descriptions of matter -- there are others. Furthermore, for most practical applications the microscopic strain is neither important nor measurable. The statistical mechanical problem is best defined in the thermodynamic limit of large systems where the macroscopic strain is the only concern. Once problems at this length scale are solved, the problems of finely divided inhomogeneous media can be approached from above, which is the usual approach in engineering calculations with, say, finite element methods.

The many-body theory that James and Guth formulated left them with the difficult problem of relating the molecular coordinates to the strain. For a homogeneous crystalline solid, the deformation of a unit cell coincides, on average, with the macroscopic deformation, and this immediately relates the cell axes to the macroscopic deformation. On the other hand, a simple fluid conforms to the shape of its container, so the state of strain is of no consequence. However, containers provide non-holonomic constraints on the configuration space of simple fluids: the coordinates are confined to a compact domain determined by the container.  In usual practice an elastomer is constrained by mechanical means over a portion of its surface by a set of clamps, a wheel rim, a road bed, etc., which are external constraints that deliver stress. As in the case of fluids, these are non-holonomic constraints that impose boundary conditions on the configuration space; for an elastomer these are invariably discontinuous boundaries. However, the very complicated constraints that might be encountered in a real application should not be solved at the level of statistical mechanics. We have to be content to evaluate the equation of state for a simple geometrical shape and leave complex geometries to engineering calculations.

In their 1943 paper\cite{11}, James and Guth wrote that ``Rubber resembles a \emph{gas} very strikingly in its thermoelastic behavior." Given this analogy, it may have been natural to think about integrating over a configuration space with boundary constraints similar to those provided by a container. James and Guth introduced holonomic constraints by picking junction points in the network that were declared to be fixed by external forces, and which are displaced by the macroscopic strain. This artifice enabled them to sidestep difficult integrations. They showed that the average coordinates of the free junctions are linear functions of the coordinates of the fixed junctions; integrations over their fluctuating positions have the majority of their support over molecular dimensions. The difficult problem of integrating all coordinates over the volume of the elastomer, subject to constraints, was thereby eliminated. 

Unfortunately, this construction merely shifted the problem to a different arena.  As Flory pointed out,\cite{14} the value that one obtains for the modulus in this treatment depends on the number and location of the fixed points. One cannot make a conclusive theory based on this treatment of constraints without additional assumptions. The rigor that is inherent in the many-body theory was vitiated by fixing junctions.

What is needed is a trick to introduce holonomic constraints which convey the dimensional information of non-holonomic constraints, but which do not require special treatment for any particular atomic species. The symmetric gyration tensor
\begin{equation}\label{eq6}
S^{ab} =N^{-1}\sum^N_{i=1}x^a_ix^b_i\sim V^{-1}\int_Vr^ar^bdV\quad 1\le a,b \le 3
\end{equation}
provides just what is needed. The continuum form on the right, with integration over the volume $V$ of the material, approximates the atomic coordinate form on the left all the better the larger (thermodynamic limit) is the system. In both the atomic and continuum forms, the coordinates are measured from the center of symmetry of the system. The macroscopic strain is defined by the deformation tensor with components $\lambda^c_a$ via
\begin{equation}\label{eq7}
S^{cd} =\lambda^c_a\lambda^d_bS^{ab}_0 \Longrightarrow {\bf S}={\bf \lambda' S_0\lambda}
\end{equation}
Here ${\bf S_0}$ is a reference state, the summation convention is used, and a matrix representation is 
conveyed by the boldface equation on the right. It is worthy of note that the equality between atomic and continuum evaluations of the gyration tensor in eq. (7) accommodates inhomogeneous media as well as homogeneous. The macroscopic strain tensor $\lambda$ defined in eq. (8) is insensitive to atomic detail; it is defined by this equation.  Note that this is a macroscopic deformation, whereas $\varepsilon$ in eq. (6) is microscopic, \emph{i.e.}, is a function of coordinates (location) in general.

\subsection*{Eckart Coordinates}
The $3\times\mu$ matrix $X=(x^a_i),1\le a\le3,1\le i\le\mu$, of cross-link coordinates gives the gyration tensor in eq. (7) as ${\bf S}=\mu^{-1}XX'$. The diagonal components of this tensor are equivalent to a multi-dimensional radius in the configuration space, as a conversion to polar form will demonstrate.  This transformation from Cartesian coordinates $X$ to polar coordinates, \emph{i.e.}, the polar decomposition of the matrix $X$, was first executed in the context of a physical problem by Eckart,\cite{30} and was first applied to polymer configuration problems by {\v S}olc.\cite{31} The transformation is the same as the well-known $SVD$ decomposition in multivariate statistics. For our needs, the transformation of the volume element is most easily carried out with use of some matrix algebra. The Autonne-Eckart-Young theorem\cite{32} enables us to write $X = R'\xi U$ where $R\in SO(3)$ is a rotation matrix in the special orthogonal (Lie) group ($RR'=R'R=1_3$), $\xi$ is a diagonal matrix such that $\xi^2$ is the matrix of eigenvalues of the non-singular matrix $XX'$, and $UU'=1_3,U'U \ne1_{\mu}$ is a point in a Stiefel manifold,\cite{33} \emph{i.e.}, $U$ lies in the coset space $O(\mu)/O(\mu-3)$. The transformation of the $3\mu$-dimensional volume element: $dX\to J[dX/d(R\xi U)]dRd\xi dU$, can be done in a few different ways. The Jacobian $J[dX/d(R\xi U)]$ can be computed directly,\cite{34} but fewer computations are required if use is made of the rule from differential geometry that $\sqrt{g}dx$ is the volume element associated to the metric $ds^2 = g_{ij}dx_idx_j$ . The Jacobian is computed in the Appendix.
\subsection*{Eliminating the Zero Eigenvalue of the Laplacian}
Since the coordinates in eq. (7) are measured relative to the origin at the center of symmetry, the rows of $X$ sum to zero. Define the orthogonal matrix $T\in SO(\mu)$ that diagonalizes the Laplacian via a similarity transformation. That is, make the substitution $X = QT$ , such that $\rm{tr}(XKX') = \rm{tr}(QTKT'Q') = \rm{tr}(Q\kappa Q')$, where $\kappa$ is the matrix of non-vanishing eigenvalues of $K$, and $Q$ is a normal coordinate representation of the configuration space. The zero eigenvalue of $K$ is generated by the constant row of $T$ , and the corresponding column in $T'$ gives a zero when it multiplies $X$. This coordinate is the center of symmetry that is fixed at the origin, and $XT'$ annihilates the corresponding column of $Q$.  One gets to the same place by deleting the constant column of $T'$, such that $T'\to T'_0$; the zeros in $Q$ and $\kappa$ combine and may be deleted, so that $Q$ becomes a $3\times(\mu - 1)$ matrix and $\kappa$ is a $(\mu - 1)\times(\mu - 1)$ positive definite diagonal matrix. Since $XX'=QT_0T'_0Q'=QQ'$ (note that $T_0T'_0=1_{\mu-1}$ but $T'_0T_0\ne1_{\mu-1}$).  The $Q$ coordinates can be reduced to polar form as described above for the cartesian matrix $X$, to give
\begin{equation}\label{eq8}
\beta \bar V =\gamma\mu\rm{tr}(SU\kappa U') 
\end{equation}
where $U$ is a $3\times (\mu-1)$ matrix representation of the coset space $O(\mu-1)/O(\mu-4)$. The dependence on the rigid body rotation $R$ cancels in eq. (9) because the trace is invariant to cyclic permutation of the arguments.  

The volume element on the configuration space begins as $dX = \delta(x_0)\prod_{i,a}dx^a_i$, where the delta function suppresses integration over the center of symmetry coordinate $x_0$. The Jacobian determinant of the transformation to normal coordinates is unity, and since the center of symmetry coordinate has been suppressed, we have $\delta(x_0)dX \sim dQ$. 

The transformation to polar coordinates (see Appendix) gives
\begin{equation*}
dX=dRdU[{\rm det}(S)]^{(\mu-5)/2}\prod_{a<b}|S_a-S_b|\prod_adS_a
\end{equation*}
to within an uninteresting constant.  The integration over the space $R\in SO(3)$ gives another constant, and  since we are to integrate over $dX/d{\bf S}$ by the prescription of eq. (2), that is, the components of the diagonal tensor $S$ are fixed, all that remains is the integral over $U$.  Happily, the structure of $\kappa$ makes this part of the problem easier than one might expect. 

\section*{Evaluation of the Configuration Integral}
The probability that the James-Guth model elastomer is found in a state $ S = {\rm diag}(S_a)$ is
\begin{equation}\label{eq9}
P(S)dS\propto [{\rm det}(S)]^{(\mu-5)/2}\prod_{a<b}|S_a-S_b| dS\int_{UU'=1} {\rm etr}(-\gamma\mu SU\kappa U')dU
\end{equation}
where $\rm{etr}(\cdot) = \exp[\rm{tr}(\cdot)]$. Terms that would convert the proportionality to an equality will cancel, as does the free energy $A_0$, when the ratio $P({\bf S})d{\bf S}/P({\bf S}_0 )d{\bf S}_0 = {\rm det}(\lambda\lambda')P({\bf S})/P({\bf S}_0)$ is computed. The integration over $dU$ covers the coset space $U\in O(\mu-1)/O(\mu-4)$. The elements of $U$ are bounded: $0\le |u^a_j|\le 1$ because $UU' = 1_3$. (The extremely interesting geometrical fact that emerges from this coordinate transformation is this: In a configuration space of $3N$ variables, three are deleted for center of mass motion, three for rigid body rotations, and three fix the macroscopic size of the system. The remaining $3N- 9$ variables comprise a \emph{non-Euclidean compact} space, independent of any model potential.)

The evaluation of the integral is highly dependent on the eigenvalue spectrum, $\kappa$, of the Laplacian. Several facts about the spectrum can be deduced from general theorems on matrices, but it is the small eigenvalues that are important. Regardless of the exact values that the small eigenvalues take, one knows that the dominant contribution to the integral in eq. (10) will be attained in the region where $U\kappa U'$ has its minimum value.  This means that the dominant contribution to the integral will be found in the region of the $U$-space that is associated with the smallest eigenvalues of $\kappa$.  In addition, the free energy must be an extensive thermodynamic function if the theory is to make any sense. But $S =O(\mu^{2/3})$ because of eq. (7). This requires that $U\kappa U' \approx O(\mu^{-2/3})$, so as to cancel the macroscopic size dependence in $S$. These observations were made long ago;\cite{35} the main purpose of the next section is to show that this estimate of the small eigenvalue is correct.

Let the eigenvalues be ordered such that $0<\kappa_1\le\kappa_2\le \cdots \le \kappa_{\mu-1}$. The minimum of ${\rm tr}(U\kappa U')$ will be attained on the subspace where $U =(R,0); R\in O(3)$ and $0$ is a $3\times (\mu-4)$ matrix of zeroes. In the next section we will see that the small eigenvalues are three fold degenerate, $\kappa_1=\kappa_2=\kappa_3$, for a cubical elastomer, which collapses the integral to
\begin{equation}\label{eq10}
\int{\rm etr}(-\gamma\mu SU\kappa U')dU \approx {\rm etr}(-\gamma\mu\kappa_1S)=\exp[-\gamma\mu\kappa_1 (S_1 +S_2 +S_3 )]
\end{equation}
This may be considered to be the leading term in an expansion; it is difficult to improve upon it.\cite{Wei}

On evaluating the contribution to the free energy from eq. (11) we get the term $\hat A=kT\mu\gamma\kappa_1(S_1 +S_2 +S_3)$, where $\gamma\kappa_1$ is the only term that depends on constraints operating at the time of formation of the elastomer.  These constraints are imprinted on the spectrum of eigenvalues of the Laplacian $K$.  The BFE average now comes to the rescue: the average over the probability distribution of constraints is a simple average over connectivity. That is, $\langle\hat A\rangle = kT\mu\langle\gamma\kappa_1\rangle(S_1 + S_2 + S_3 )$ requires a straightforward statistical mechanical average of the small eigenvalue.  It is hard to imagine a technique for evaluating the small eigenvalues that does not invoke a BFE averaging procedure. Recognition that $\gamma\kappa_1$ is a BFE average, even with brackets omitted, is implicit in everything that follows.  If the network is formed with a distribution of chain lengths (MWD) the averaging of the small eigenvalue requires considerable care, and is left for another time.

\section*{Estimating the Small Eigenvalues}
\emph{The ``. . . smallest eigenvalue of the Laplacian, is far from trivial; in fact, it is difficult to overemphasize its importance."}\cite{22}

The smallest non-zero eigenvalue has been called the ``algebraic connectivity" of a graph.\cite{36} Unfortunately, what is known about $\kappa_1$ in random graph theory literature is not very helpful; what sets physical random graphs apart from mathematicians' random graphs\cite{37} is that physical random graphs are embedding in $\mathbb R^3$ with a more or less uniform density of vertices, and that has a profound influence on the spectrum at the small eigenvalue edge. For our purposes an embedded graph in a roughly isometric hard or soft solid has a number density $\rho = \mu/V$ of vertices such that any slice through the graph with thickness $\ell= \rho^{-1/3}$ and area $A={\rm O}(V^{2/3})$ contains ${\rm O}(\mu^{2/3})$ vertices. For mathematical purposes this asserts a uniformity of density of vertices (to within the usual influence of the discontinuity at the surface) while allowing for short-range correlations that may be of interest (and will be of great interest for many problems).

Using this notion of slices with thickness $\ell$, a cube with volume $\ell^3$ contains one vertex on average. The connections between a vertex and its neighbors depend, of course, on the detailed edge-length distribution in the graph. Suppose that a cubical elastic body is divided into a simple cubic lattice of $\mu=n^3$ cells. A layer of these cells contains, on average, $n^2$ cross-links. The number distribution of cross-links in every layer will be a Poisson distribution, with variance $n$, if the cross-links are distributed at random. Since we are interested in the large $\mu$ behavior, the fluctuations in the number of vertices in the layers is sufficiently small to be neglected [variance/mean = O($ n^{-1}$)]. Because these layers contain large numbers of vertices the statement effectively ignores short-range correlations. This is not an assertion that short-range correlations are unimportant. 

Our eigenvalue problem now inherits an enumeration scheme that looks like that used to calculate the vibrations in a simple cubic crystal. The elastomer does not have phonon modes in the sense of a crystalline solid.  (However, the long wavelength modes are visible in a soft material as macroscopic oscillations or vibrations.)  Instead, we have a connectivity problem that is mathematically analogous to that of a crystalline solid. The long wavelength eigenvectors of the Laplacian matrix are the same as the long wavelength eigenvectors in a crystalline solid. It is well known that the long wavelength modes in a crystal determine the lowest energy excitations, and the long wavelength eigenvectors, with wavelength $\lambda \gg {\rm max}(\ell, \langle r^2\rangle^{1/2})$, will generate the small eigenvalues of our graphs. James\cite{13} calculated some statistical properties of a regular cubic lattice in relation to his work on elasticity, so this idea is not new. What is new is a better way to handle the randomness that overrides the lattice.  The fundamental reason for discretizing the space of the soft material is to borrow the matrix indexing scheme from crystal theory, where there is a one-to-one correspondence between coordinates and matrix indices; this greatly facilitates the construction of eigenvectors.  There is no obvious indexing of a random graph, so the discretization is imposed to provide this correspondence. 

Index the cells in our subdivided cubical elastomer in the obvious way as ${\bf j}=\{j_1, j_2, j_3\}$, where $1\le j_a \le n$. The probability $p(r)d{\bf r}\approx p({\bf j}^{\alpha},{\bf j}^{\beta})\ell^3$ that two average cells with multi-indices ${\bf j}^{\delta}$ are connected by a polymer chain is discretized as
\begin{equation}\label{eq11}
p({\bf j}^{\alpha},{\bf j}^{\beta})\ell^3 =\bar f(\gamma_*/\pi)^{3/2}\exp[-\gamma_*\ell^2({\bf j}^{\alpha}-{\bf j}^{\beta})^2]\ell^3.
\end{equation}
Summed over all ${\bf j}^{\beta}$, eq. (12) yields the average number, $\bar f$, of chains connected to the cross-link located at ${\bf j}^{\alpha}$.  We are assuming a homogeneous elastomer, such that the number of chains connected to a cross-link selected at random is translationally invariant.  In eq. (12) the parameter $\gamma_*=3/2\langle r^2\rangle_*$ is determined by the chain statistics at the time the network is formed.  The fundamental polymer physics that goes into the model elastomer asserts that $\langle r^2\rangle_*$ is the unperturbed mean-square end-to-end distance of a free chain at the temperature of cure.  This is a fixed length parameter that will not vary with temperature.  At the temperature of measurement, where we are computing the configuration integral in eq. (11), the corresponding parameter governing the potential of mean force is temperature dependent.  At the time the network is formed the only role for the chain length distribution is to determine the connectivity.  After the network is formed the chains assume their role of delivering stresses. 

The long wavelength (unnormalized) eigenvectors for a simple cubic lattice has components of the form
\begin{equation*}
\exp[\pi i(j_1m_1+j_2m_2 +j_3m_3)/n]
\end{equation*}
where $1\le j_a\le n$; $0\le m_a < m_e, 1\le a\le 3$, and $i=\sqrt{-1}$. The index set ${\bf m}=\{m_1,m_2,m_3\}$ labels the eigenvalues $\kappa_{\bf m}$.  The restriction $m_e\ll n$ limits the eigenvectors to long wavelengths; if the wavelength is less than several multiples of $\langle r^2\rangle^{1/2}_*$ the local structure of the graph will become important, and the corresponding eigenvectors will be different from simple Fourier functions (although one could write the exact  eigenvectors as linear combinations of Fourier functions since the latter form a complete basis). The constant eigenvector, with $m_1 = m_2 = m_3 = 0$, generates the zero eigenvalue of $K$. The components of the eigenvectors for the simple cubic lattice are actually $\prod_a\cos[( j_a-1/2)m_a\pi/n]$. The approximate complex version makes subsequent calculations simpler and does no damage. We have no need of toroidal boundary conditions, which would insert a factor of 2 in the trigonometric functions and artificially render the eigenvalues doubly degenerate.  

After going to the trouble of discretizing the space we will now undo this work by replacing sums by integrals. Of course, the motivation for discretizing is to make the physical picture clear and to help guide an understanding of the implications of the long wavelength regime where the method will be valid. Our eigenvector is now specialized to a particular plane wave for which $m_1 = 1$ and $m_2 = m_3 = 0$.   We could just as well have chosen $m_2$ or $m_3$ to be the only non-zero index. That is, the isotropy of the network for a cubical elastomer renders the smallest non-zero eigenvalue triply degenerate. Since the selected eigenvector is constant on planes perpendicular to the $x$-axis, it picks up an entire layer of vertices in the slab of thickness $\ell$, and thereby averages the connectivity over the slab.  Our eigenvalue problem is now mapped into
\begin{align}\label{eq12}
\kappa_1t_1=&{}Kt_1\nonumber\\
\kappa_1\exp(\pi i x_1/n\ell)=&{}\bar f\{\exp(\pi i x_1/n\ell)-(\gamma_*/\pi)^{1/2}\int\exp[-\gamma_*(x_1-x_2)^2]\exp(\pi i x_2/n\ell)dx_2\\
\kappa_1=&{}\bar f\{1-(\gamma_*/\pi)^{1/2}\int\exp[-\gamma_*(x_1-x_2)^2]\exp[-\pi i (x_1-x_2)/n\ell]dx_2\}\nonumber\\
=&{}\bar f\{1-\exp[-\pi^2/(4\gamma_*n^2\ell^2)]\}\nonumber\
\end{align}
(There are cancelled factors of $n^2$ that arise from summing over the constant $x_2$ and $x_3$ components of the eigenvectors on the left and over the cross-links in the slices perpendicular to the $x_1$-axis on the right.) The first factor in brackets comes from diagonal elements of the matrix, and the second factor, the integral, arises from the average number $\bar f$ of off-diagonal elements, each of which has the value $-1$ in the Laplacian (the actual value will be $-k$, where $k$ is the number of chains that connect the two cross-links in question to allow for multiple connectivity). 

Since $n$ is as large as we like, it follows that the first several small eigenvalues will be given by
\begin{equation}\label{eq13}
\kappa_{\bf m}=\bar f\{1-\exp[-\pi^2{\bf m}^2/(4\gamma_*n^2\ell^2)]\}\approx \bar f \pi^2{\bf m}^2/(4\gamma_*V_*^{2/3})= \pi^2{\bf m}^2\bar f\langle r^2\rangle_*/(6V_*^{2/3})
\end{equation}
where ${\bf m}^2 = m^2_1 + m^2_2 + m^2_3; |{\bf m}| < m_e$ . We now have the needed proof that the small eigenvalues are proportional to $\mu^{-2/3}$. It is also clear that in the limit as $\kappa\to 0_+$, the spectral density $g(\kappa)$ of small eigenvalues of the Laplacian matrix tends to the same distribution that Debye calculated for a spherical continuum elastic solid. The only difference is that our long wavelength density is determined by microscopic parameters -- the chemistry in $\bar f$ and chain statistics in $\gamma_*$ -- whereas Debye's spectrum is determined by phenomenological Lam\'e constants. Standard textbooks\cite{38} explain the relation between the long wavelength spectrum for the Born-von Karman model of a simple cubic lattice and the Debye low frequency spectrum.

If there is a distribution of molecular weights (MWD) of the chains connecting cross-links the average over this distribution couples with the graph eigenvalue spectrum, requiring a very careful analysis of the averaging procedure.  In first approximation this MWD average can be postponed to eq. (14), where it simply gives an average of $\gamma^{-1}_*\sim \langle r^2\rangle_*$.  

\section*{Summary}
On combining eqs. (1,3,11, and 14), and replacing the probability distribution by the equivalent Helmholtz free energy, one obtains
\begin{align}\label{eq14}
A(S)=&{}A_0(|S|)+[kT\mu\bar f\pi^2\langle r^2\rangle_*/(4\langle r^2\rangle/V_*^{2/3})](S_1+S_2+S_3)\nonumber\\
&{}-(\mu-5)kT\ln(|S|^{1/2})-kT\ln(|S_1-S_2||S_2-S_3||S_1-S_3|)\
\end{align}
Now that this result is at hand one can see that the last term in eq. (15), which causes the eigenvalues of $X$ to repel one another, is not of thermodynamic significance. It can contribute, say, a surface tension
term only if at least one of the arguments $|S_a-S_b| = {\rm O}[\exp(\mu^{-2/3})]$. But this is ridiculously small; normal thermodynamic fluctuations in macroscopic lengths will be ${\rm O}(\mu^{-1/3})$, so that the term is at most ${\rm O}[\ln(\mu)]$, and may be neglected. Furthermore, one can ignore the difference between $\mu$ and $\mu-5$ in the preceeding logarithmic term.

The next point to make is that the theory gives a stable un-stressed state. This state is a
minimum of the free energy; that is
\begin{equation*}
\partial A(S)/\partial S_a= \frac{\partial A_0(|S|)}{\partial |S|}\frac{\partial (|S|)}{\partial S_a}+G-\mu kT/2S_a=G-\frac{\mu kT/2-(\partial A_0(|S|)/\partial \ln(|S|)}{S_a}
\end{equation*}
vanishes when evaluated at $S_a =S^0_a$. Here $G= kT\mu\bar f\pi^2\langle r^2\rangle_*/(4\langle r^2\rangle V_*^{2/3})$.  At the minimum 
\begin{equation*}
S^0_a=\frac{\mu kT/2-(\partial A_0(|S|)/\partial \ln(|S|)_{S=S^0}}{G}=\frac{\mu kT/2+p_0V/2}{G}
\end{equation*}
with the last version resulting from $|S|\propto V^2$.  

The term involving $p_0$ can be discarded with the following argument:  Suppose we had a good theory for $A_0$, such that one could solve for the equilibrium volume $(\partial A_0/\partial V)_T=-p_0=0$: the unstrained (uncompressed) state of the base polymer is a minimum when the hydrostatic pressure vanishes. The difference between the base polymer free energy and the approximate $A_0$ that has been factored from the configuration integral in eq. (3) is owing to the constraints of the cross-links that are implicit in $A_0$, and this difference can be made small by decreasing the density of cross-links. The term with $p_0$ may be set to zero, so that the only volume dependence that survives is that in $G$, so that $S^0_a ={\rm O}(V^{2/3})$ as it must.  What is most important about this observation is that \emph {the James-Guth theory does not have a physically realistic un-stressed state in the absence of the $\log(V)$ term}.  

The gyration tensor in the unstrained state of our cubical elastomer has equal diagonal components
\begin{equation*}
S^0_a=V^{-1}\int^{L_0/2}_{-L_0/2}x^2_adx_1dx_2dx_3=L^2_0/12
\end{equation*}
giving $S_a =\lambda^2_a(L^2_0/12)=\lambda^2_aV^{2/3}_0/12$. It also follows that $|S|/|S^0| =(V/V_0)^2$. Putting these pieces into eq.(15) gives
\begin{equation}\label{eq15}
\Delta A= \Delta A_0+[kT\mu(\frac{\bar f\pi^2}{4\cdot 12})(\frac{\langle r^2\rangle_*}{\langle r^2\rangle})(\frac{V_0}{ V_*})^{2/3}(\lambda^2_1+\lambda^2_2+\lambda^2_3-3)-\mu kT\ln(\frac{V_0}{V}).
\end{equation}
The ratio $\langle r^2\rangle_*/V_*^{2/3}$ came from the eigenvalue spectrum and is fixed at the time of cure; this ratio is independent of temperature. Measurement of the stress-temperature coefficient gives $d\langle r^2\rangle/dT$, not $d\langle r^2\rangle_*/dT$. (This nomenclature differs from that used in other work so as to make clear that there are three states involved -- the state of cure at $T_*$, the unstressed state at temperature $T$, and the stressed state at temperature $T'$.)  The number of chains in the network is $\nu = \mu\bar f/2$, and if one ignores the (usually very small) differences between $V_*, V, {\rm and} V_0$, the elastic free energy simplifies to 
\begin{equation*}
\Delta A_{el}=\nu kT(\pi^2/24)(\frac{\langle r^2\rangle_*}{\langle r^2\rangle})(\lambda^2_1+\lambda^2_2+\lambda^2_3-3).
\end{equation*}
Note that $\bar f$ differs from the maximum chemical functionality so that it effectively corrects for dangling chains and loops, and therefore $\nu= \mu\bar f/2$ is the count of ``elastically effective" chains.  The surprising presence of $\pi$ in the modulus (front factor) is a consequence of the estimate of the small eigenvalue of the Laplacian that was made with use of a plane wave eigenvector. Of course, $\pi$ appears naturally in the Rouse spectrum of relaxation times; it should not be considered unusual that it arises in the bulk elastomer context.  Note that $\pi^2/24\approx 0.41$, which is is comparable to values for this factor that have appeared in the work by James and Guth and the later work of Flory invoking the cycle rank.

\section*{Conclusion}
It is remarkable that the James-Guth many-body theory of elasticity gives essentially the same result that Flory deduced over a period of several decades. The logarithmic term, chain dimension ratio, and front factor ($\pi^2/24$ rather than the cycle rank\cite{14}) all emerge when the problem is handled in a natural way. As this work shows, what is natural requires some fairly intense mathematics (at least relative to the simpler Wall-Flory treatment), but the reward is a theory of elasticity that has a firm foundation. Very few pproximations have been made along the way; the most critical (and the hardest) place to make improvements is in the estimate of the small eigenvalue of the Laplacian of the graph. The integral in eq. (11) might also be improved\cite{39} with deeper understanding of the spectrum. Improvements in the potential alter the model, which goes beyond the present concern.

A final comment regarding potential improvements to the model potential needs to be made. If the many-body potential of mean force is to be augmented, for example with constraints,\cite{9} these must be inserted at the level of the coset space. That is, constraints depend on $U-U_\otimes$, where $U_\otimes$ is the locus of constraints, and not on $X-X_\otimes$.  This is a critical observation because: (i) microscopic constraints cannot alter the macroscopic dimensions directly, although they have an indirect influence through the strength of microscopic forces, and (ii) the logarithmic term has to be maintained in the form of $const.\times \mu\ln(V)$ so as to yield a stable un-stressed state.  

\section*{Acknowledgement}
The author is grateful for the hospitality of Profs. U. Suter and P. G\"unter at the ETH in Zurich during the spring of 2008 when a portion of this work was completed.

\subsection*{Appendix: Polar Coordinates for Matrices}
It seems that a heavy calculation is unavoidable to get the volume element for polar coordinates of matrices.  The approach used here makes use of the well-known relation between the metric $ds^2=g_{ij}dx_idx_j, 1\le i,j\le n$  for an $n$-dimensional real space and the volume element $\sqrt{g}d{\bf x}$, where $d{\bf x}=\prod_idx_i$.  If coordinates are arranged in an $m\times n, m\le n$ real matrix $X$ the volume element is $\prod_{a,j}dx_{aj}, 1\le a\le m, 1\le j\le n$, and clearly this belongs to the metric $ds^2={\rm tr}(dXdX')$ .  (Note that the rank of $X$ is assumed to be $m$ throughout this discussion.)   The differential of the polar decomposition,\cite{32,33}, $X=R'\xi U$, is
\begin{equation}\label{a1}
dX=R'(\delta r\xi U+d\xi U+\xi dU)
\end{equation}
where $\delta r=RdR'$ is a skew-symmetric matrix.  Here $R\in O(m)$ spans the group of orthogonal matrices, $\xi^2$ are the eigenvalues of $XX'$, and $UU'=1_m$, but $U'U\ne 1_n$ if $m\ne n$; in the latter case $U$ is a Stiefel manifold, {\emph i.e.}, the coset space $U\in O(n)/O(n-m)$, and if $m=n, U\in O(n)$.  Inserting eq. (17) into the metric and simplifying gives 
\begin{equation}\label{a2}
ds^2={\rm tr}[d\xi^2+(\delta r\xi+\xi\delta u)(\xi\delta r'+\delta u'\xi)+\xi^2dU(1-U'U)dU']
\end{equation}
Similar to $\delta r$, $\delta u = UdU'$ is skew-symmetric.  If $m=n$ the last term vanishes.

The middle term in eq. (18) is simplified by extracting symmetric and skew-symmetric parts with $\delta r\xi+\xi\delta u=(\xi\delta\phi+\delta\phi'\xi)+(\xi\delta\theta-\delta \theta'\xi)$, where $\delta\phi=(\delta u -\delta r)/2$ and $\delta\theta=(\delta u +\delta r)/2$ are both skew-symmetric.  But now the first term in parentheses is symmetric and the second is skew-symmetric, which enables one to simplify the expression ${\rm tr}[(s+a)(s'+a')]={\rm tr}[(s+a)(s-a)]={\rm tr}(s^2-a^2)$ where $s$ and $a$ are the symmetric and skew-symmetric, respectively.  In component form, this term becomes
\begin{equation*}
{\rm tr}(\delta r\xi+\xi\delta u)(\xi\delta r'+\delta u'\xi)=2\sum_{a<b}[(\xi_a-\xi_b)^2\delta\phi^2_{ab}+(\xi_a+\xi_b)^2\delta\theta^2_{ab}]
\end{equation*}
with associated volume element 
\begin{equation*}
\prod_{a<b}|\xi^2_a-\xi^2_b|d\phi_{ab}d\theta_{ab}\sim \prod_{a<b}|\xi^2_a-\xi^2_b|dr_{ab}du_{ab},
\end{equation*}
where numerical factors of no consequence to thermodynamics have been dropped. 

The last term in eq. (18) requires some work, but is instructive.  Write a partitioned $U$ as $U=(U_1,U_2)=U_1(1,y)$, where $y=U^{-1}_1U_2$.  The matrix $U_1$ is $m\times m$ and is non-singular except on subspaces of lower dimension.  Since $UU'=1$ it follows that $UU'=U_1(1+yy')U'_1\to 1+yy'=(U'_1U_1)^{-1}$.  Some algebra establishes
\begin{equation*}
1-U'U=\left({\begin{array}{*{20}c}
y\\
-1\\
\end{array}}\right)(1+y'y)^{-1}(y', -1)=(y,-1)_c(1+y'y)^{-1}(y', -1)
\end{equation*}
with use of $y'(1+yy')=(1+y'y)y'$ and its variations. The dimension of the $y$-manifold, a Grassmannian, is $m(n-m)$.  Since $dU = (dU_1, dU_1y+U_1dy)$ it follows that $dU(y,-1)_c=-U_1dy$, so that 
\begin{equation}\label{a3}
{\rm tr}|[\xi^2dU(1-U'U)dU']={\rm tr}|[U'_1\xi^2U_1dy(1+y'y)^{-1}dy']
\end{equation}
The final detail that one needs to calculate the volume elements is this: Let $dZ=(dz_{aj}), 1\le a\le m, 1\le j\le n$, be an $m\times n$ matrix with row form 
\begin{equation*}
dz=(dz_{11},dz_{12}, \cdots, dz_{1n}, dz_{21}, \cdots, dz_{2n}, \cdots, dz_{mn}); 
\end{equation*}
it can be seen that ${\rm tr}(AdZBdZ')=dz(A'\otimes B)dz'$, where $A$ and $B$ are conformable. The volume element associated with this metric is $dV= |A|^{n/2}|B|^{m/2}\prod dz_{aj}$.  The volume element associated to the component defined in eq. (19) is 
\begin{align*}
|U'_1\xi^2U_1|^{(n-m)/2}|1+y'y|^{-m/2}|\prod dy_aj=&{}|\xi^2|^{(n-m)/2}|1+yy'|^{-(n-m)/2}|1+y'y|^{-m/2}\prod dy_{aj}\\
=&{}|\xi^2|^{(n-m)/2}|1+yy'|^{-n/2}\prod dy_{aj}
\end{align*}
Putting the pieces together one finds
\begin{align}\label{a4}
\prod_{a,j}dx_{aj}=&{}{\rm const.}\{|\xi^2|^{(n-m)/2}\prod_{a<b}|\xi^2_a-\xi^2_b|\prod_ad\xi_a\}\{drdu\}\{|1+yy'|^{-n/2}\prod^m_{a=1}\prod^{n-m}_{k=1}dy_{ak}\nonumber\\
=&{}{\rm const.}\{|\xi^2|^{(n-m)/2}\prod_{a<b}|\xi^2_a-\xi^2_b|\prod_ad\xi_a\}dRdU\
\end{align}
where $dR$ and $dU$  are short-hand notations for the volume elements on $SO(m)$ and $SO(n)/SO(n-m)$, respectively.  A check of physical dimensions ($L=\rm{length}$) will show that ${\rm dim}(dX)={\rm dim}[|\xi^2|^{(n-m)/2}\prod_{a<b}|\xi^2_a-\xi^2_b|\prod_ad\xi_a]\sim L^{mn}$, since the orthogonal group and the Steifel manifold are dimensionless. 


\begin{thebibliography}{99}

\bibitem{1}
Flory, P. J.
\emph{Principles of Polymer Chemistry}; 
Cornell University Press: Ithaca, NY, 1953.
\bibitem{2}
Kuhn, W.,
\emph{Kolloid Zeits.} 1936, 76, 258.
\bibitem{3} 
Wall, F. T. \emph{J. Chem. Phys.} 1942, 10, 485.
\bibitem{4} 
Wall, F. T. \emph{J. Chem. Phys.} 1942, 10, 132.
\bibitem{5} 
Flory, P. J.; Rehner, J., Jr. \emph{J. Chem. Phys.} 1943, 11, 512.
\bibitem{6} 
Flory, P. J.; Rehner, J., Jr.\emph{ J. Chem. Phys.} 1943, 11, 521.
\bibitem{7} 
Treloar, L. R. G. \emph{The Physics of Rubber Elasticity, 3 ed.}; Clarendon Press: Oxford, 1975.
\bibitem{8} 
Mark, J. E.; Erman, B. \emph{Rubberlike Elasticity: A Molecular Primer}; Wiley: New York, 1988.
\bibitem{9} 
Erman, B.; Mark, J. E. \emph{Structures and Properties of Rubberlike Networks}; Oxford: New York,
1997.
\bibitem{10} 
Graessley, W. W. \emph{Polymeric Liquids and Networks: Structure and Properties}; Garland Scientific: New York, 2004.
\bibitem{11}
 James, H. M.; Guth, E. \emph{J. Chem. Phys.} 1943, 11, 455.
\bibitem{12} 
James, H. M.; Guth, E. \emph{J. Chem. Phys.} 1947, 15, 669.
\bibitem{13} 
James, H. M. \emph{J. Chem. Phys.} 1947, 15, 651.
\bibitem{14} 
Flory, P. J. \emph{Proc. R. Soc. Lond. A} 1976, 351, 351.
\bibitem{15} 
Flory, P. J. \emph{J. Chem. Phys.} 1950, 18, 108.
\bibitem{16} 
Flory, P. J. \emph{J. Am. Chem. Soc.} 1956, 78, 5222.
\bibitem{17} 
Flory, P. J.; Hoeve, C. A. J.; Ciferri, A. \emph{J. Polym. Sci.} 1959, 34, 337.
\bibitem{18}
 Eichinger, B. E. \emph{Theory of Elasticity}. In \emph{Annual Review of Physical Chemistry}; Rabinovitch, B. S., Schurr, J. M., Strauss, H. L., Eds.; Annual Reviews, Inc.: Palo Alto, CA, 1983; Vol. 34; pp 359.
\bibitem{19} 
Mark, J. E. \emph{J. Phys. Chem.} B 2003, 107, 903.
\bibitem{22}
 Bollabas, B. \emph{Modern Graph Theory}; Springer: New York, 1998; Vol. 184.
 \bibitem{6a}
Brotzman, R. W., Jr.; Eichinger, B. E., \emph{Macromolecules}1983, 16, 1131.
\bibitem{6b}
Neuburger, N. A.; Eichinger, B. E., \emph{Macromolecules}1988, 21, 3060.
\bibitem{6c}
Zhao, Y.; Eichinger, B. E., \emph{Macromolecules}1992, 25, 6996.
\bibitem{23}
 Eichinger, B. E. \emph{Macromolecules} 1972, 5, 496.
\bibitem{24} 
Eichinger, B. E. \emph{Macromolecules} 1980, 13, 1.
\bibitem{25} 
Brout, R. \emph{Phys. Rev.} 1959, 115, 824.
\bibitem{26} 
Fixman, M.; Zeroka, D. \emph{J. Chem. Phys.} 1968, 48, 5223.
\bibitem{27} 
Eichinger, B. E.; Fixman, M. \emph{Biopolymers} 1970, 9, 205.
\bibitem{28} 
Edwards, S. F.; Freed, K. F. \emph{J. Phys. C: Solid State Phys.} 1970, 3, 760.
\bibitem{29} 
Murnaghan, F. D. \emph{Finite Deformation of an Elastic Solid}; John Wiley: New York, 1951.
\bibitem{30} 
Eckart,C.\emph{Phys.Rev.}1934,46,383.
\bibitem{31} 
Solc, K. \emph{Macromolecules} 1973, 6, 378.
\bibitem{32} 
Ben-Israel, A.; Greville, T. N. E. \emph{Generalized Inverses: Theory and Applications}; Wiley-
Interscience: New York, 1974.
\bibitem{33} 
Kobayashi, S.; Nomizu, K. \emph{Foundations of Differential Geometry. Vol. II}; Interscience: New York, 1969.
\bibitem{34} 
Mehta, M. L. \emph{Random Matrices, 2 ed.}; Academic Press: San Diego, 1991.
\bibitem{35} 
Shy, L. Y.; Eichinger, B. E. \emph{J. Chem. Phys}. 1989, 90, 5179.
\bibitem{Wei}
Wei, G. Y.; Eichinger, B. E. \emph{Ann. Inst. Statist. Math.} 1993, 45, 467.
\bibitem{36} 
de Abreu, N. M. M. \emph{Lin. Alg. Appl}. 2007, 423, 53.
\bibitem{37} 
Janson, S.; Luczak, T.; Rucinski, A. \emph{Random Graphs}; John Wiley: New York, 2000.
\bibitem{38} 
Ashcroft, N. W.; Mermin, N. D. \emph{Solid State Physics}; Saunders: Philadelphia, 1976.
\bibitem{39} 
Wei, G.; Eichinger, B. E. \emph{Macromolecules} 1990, 23, 4845.

\end{thebibliography}
\end{document}